\documentclass[conference,a4paper]{APSIPA2021}
\usepackage{multirow}
\usepackage{amsmath}
\usepackage[psamsfonts]{amssymb}
\usepackage{amsxtra}
\usepackage{threeparttable}
\usepackage{graphicx}

\begin{document}

\title{Jointly Predicting Emotion, Age, and Country Using Pre-Trained Acoustic Embedding}

\author{%
\authorblockN{%
Bagus Tris Atmaja\authorrefmark{1}, Zanjabila\authorrefmark{2}, and
Akira Sasou\authorrefmark{1}
}
\authorblockA{%
\authorrefmark{1}
National Institute of Advanced Industrial Science and Technology, Japan}

\authorblockA{%
\authorrefmark{2}
Sepuluh Nopember Institute of Technology, Indonesia}
}

\maketitle
\thispagestyle{empty}

\begin{abstract}
  In this paper, we demonstrated the benefit of using pre-trained model to extract acoustic embedding to jointly predict (multitask learning) three tasks: emotion, age, and native country. The pre-trained model was trained with wav2vec 2.0 large robust model on the speech emotion corpus. The emotion and age tasks were regression problems, while country prediction was a classification task. A single harmonic mean from three metrics was used to evaluate the performance of multitask learning. The classifier was a linear network with two independent layers and shared layers, including the output layers. This study explores multitask learning on different acoustic features (including the acoustic embedding extracted from a model trained on an affective speech dataset), seed numbers, batch sizes, and normalizations for predicting paralinguistic information from speech.
\end{abstract}

\begin{IEEEkeywords}
  speech emotion recognition, affective computing, acoustic embedding, multitask learning, age prediction, country prediction
  \end{IEEEkeywords}
  
\section{Introduction}
Research on computational paralinguistics advances with the advent of artificial intelligence, big data, and speed up internet access. Among many issues, computational paralinguistics research still relies on limited small data for experiments \cite{Batliner2020}. The need to conduct research on big data is needed for generalization and performance improvements.

A large-scale dataset was made for accelerating research on computational paralinguistics, the HUME-VB dataset \cite{Baird2022}. The dataset contains more than 36 hours of audio data with a 48 kHz of the sampling rate. The samples in the dataset were labeled with three pieces of information: intensity rating for ten different expressed emotions (float in ranges [0, 1]), age (integer), and country (string). This dataset was created to support the following tasks: multitask learning, emotion generation, and few-shot learning. Multitask learning includes emotion recognition from vocalizations.

% speech emotion recognition
Speech emotion recognition is a branch of computational paralinguistics that deal with the accurate prediction of emotional score/class from speech. Since the expression of speakers' emotions can be perceived by the human ears of listeners, it is also possible for the computer to have the same ability: recognize emotion from the sound. Instead of speech, which has semantic or linguistic meanings, the burst of brief vocalizations is an interesting source for predicting expressed emotion in human social life since it contains rich information on emotion \cite{Cowen2019}.

% mapping
Humans communicate emotion through two different kinds of vocalizations, prosody and vocal burst \cite{Scherer1986}. Prosody interacts with the words (linguistic) to convey feelings and attitudes via speech, while vocal burst just occurs without linguistic meanings. Examples are laughs, cries, sighs, shrieks, growls, hollers, roars, and oohs. While the study of the prosody of speech for speech emotion recognition is widely conducted (indicated by the number of available datasets), the study of the vocal burst emotion recognition is currently undergoing with support of new vocalization datasets, e.g.,  \cite{ Schuller2022,Baird2022}.

% multitask learning
It is not the only emotion that can be recognized in the human's voice. Age, gender, and nationality could also be detected from voices. The change of voice in age is recognizable by both humans and computers, although it is the hardest among emotion and gender \cite{Kaya2017}. By transferring information about age and gender, the recognition of dimensional could be improved \cite{Zhao2018a}. The task of combining several tasks together based on the same or different inputs is known as multitask learning.

% ICML ExVo Competition
At the ICML Expressive Vocalizations (ExVo) Workshop and Competition, a multitask learning task is held to predict the average intensity of each of 10 emotions perceived in vocal bursts, the speaker's age, and the speaker's  Native-Country. The challenge utilized a large vocalization dataset \cite{Cowen2019} which was shared across three different tasks. The emotion and predictions are regression problems; native-country prediction is a classification problem. The participants were evaluated by a single score of the harmonic mean from these three problems.

% contribution
This paper contributes to evaluating the pre-trained speech embeddings extractor, which was trained specifically on an affective speech dataset, to jointly predict emotion, age, and the country as in the ExVo challenge. We hypothesize that using this kind of acoustic feature extractor will lead to better results than traditional feature extractors, which extract physical information of audio signals. The rest of this paper explains related work, methods, results and discussion, and conclusion.

\section{Related work}
Research on multitask learning, predicting several tasks simultaneously using unified architecture, progressively increased due to its effectiveness in predicting several outputs with the same or similar inputs. Speech processing is an ideal test bed for multitask learning. Several pieces of information could be extracted from the same speech input, whether it is text (automatic speech recognition, ASR), gender, age, nationality, emotion, language, and disease. The following is a resume of the work of multitask learning done in the past with key differences from this study presented at the end of this section.

% general multitask learning
% parthasarathy
Parthasarathy and Busso \cite{parthasarathy2017jointly} proposed two schemes of multitask learning architectures for evaluating valence, arousal, and dominance simultaneously. The first architecture is without an independent layer for each task (shared layers only), while the second architecture is with independent layers for each task. They found the second architecture with independent layers performed better than the first one without independent layers. The scores (measured in concordance correlation coefficient) also showed improvements from the baseline architecture with single-task learnings.

% lee
Lee \cite{Lee2019} also evaluated two architectures of multitask learning optimized for emotion recognition. The first architecture is composed of three task-specific softmax layers to predict gender, emotion, and language. The second architecture is composed of one softmax layer containing two tasks, language and emotion. The results revealed a better generalization of the second architecture to predict emotion categories.

Kim et al. \cite{Kim2017} proposed to use gender and naturalness information to minimize the large mismatch between the training and test data in speech emotion recognition in the wild. The method employed traditional acoustic features ($f_{o}$, voice probability, zero-crossing-rate, MFCC with their energies and first-time derivatives) extracted on frame-level and calculated the high-level features on top of it. The high-level features then were fed into an extreme learning machine to predict the categories of emotion. They obtain significant improvement over the baseline with single-task learning. The method was evaluated on five different datasets for generalization.

%  atmaja
Atmaja and Akagi \cite{Atmaja2020b} evaluated different loss functions -- CCC, MSE, and MAE -- for multitask learning dimensional emotion recognition. The traditional approach of deep learning commonly employed MSE loss to minimize the error and training stage. Since the goal is to maximize CCC, they proposed to use CCC loss as the loss function to replace MSE. The results on two datasets (IEMOCAP and MSP-IMPROV) and two acoustic features (GeMAPS and python Audio Analysis) showed the consistency that CCC loss is superior to MSE and MAE.

% li kai
Li et al. \cite{Li2020} proposed additional information on age, gender, and emotion for speaker verification using multitask learning and domain adversarial training. The multitask learning part minimizes losses of three variables (speaker, gender, and nationality). The domain adversarial training, which also employs multitask learning, minimizes losses of speaker and emotion. The results showed that multitask improved the performance from the baseline by about 16\% while the domain adversarial training improved the performance from the baseline by about 22\%. The baseline used ResNet networks.

%  cai
Cai et al. \cite{Cai2021} employed multitask learning by predicting text characters and predicting emotion in the training phase. The model is trained to minimize the loss of categorical emotion (cross-entropy) and loss of character recognition (connectionist temporal classification). The inference phase removes the character recognition path to predicting emotion categories only. The proposed multitask learning achieved an accuracy of 78\% compared to 72\% of the baseline method with capsule networks.

% atmaja
Atmaja et al. \cite{Atmaja2022a} evaluated multitask learning of emotion recognition (dimensional) naturalness scores from speech. They evaluated two different architectures with and without independent layers. The architecture without independent layers (shared layers only) exhibits the best performance in predicting valence, arousal, and dominance scores. The shared layers have been built using three layers of fully connected networks with nodes of 512, 256, and 128. However, the scores for naturalness recognition in multitask learning is lower than in single-task learning.

Instead of focusing on the single task evaluation on multitask learning (e.g., only predicting emotion in multitask emotion and transcription \cite{Cai2021}, or emotion and language \cite{Lee2019}), this study focused on the all tasks evaluated on the multitask learning. We employed the harmonic mean evaluation from three metrics for three tasks and used this harmonic mean as the final evaluation. This multitask learning evaluation using all tasks was not evaluated on the previous tasks, where the authors only focused on emotion recognition or speaker verification.

\section{Methods}
\subsection{Datasets}
This study relies on the HUME-VB dataset, which is used at the ICML Expressive Vocalization (ExVo) Competition 2022. The dataset is a large-scale emotional non-linguistic vocalization known as vocal burst. An example of this vocal burst is ``argh!'' to express distress emotion. There are ten emotions rated in continuous scores. These emotions are Amusement, Awe, Awkwardness, Distress, Excitement, Fear, Horror, Sadness, Surprise, and Triumph. The data were collected from 1702 speakers aged 20-39 years old. The collection locations are China, South Africa, Venezuela, and the US. The total duration of the dataset is almost 37 hours (36 hours 47 minutes). Although the data were split into train, validation, and test, the test set was closed by the organizer of the competition. Hence, we evaluated our methods mostly on the validation set (except in the last part, where test results are reported). More details about the dataset can be found in the \cite{Baird2022}.

\subsection{Pre-trained Acoustic Embedding}
We evaluated a pre-trained model finetuned on an affective speech dataset. The model \cite{Wagner2022,Wagner2022a} is based on wav2vec2-large-robust model \cite{Hsu2021a}. The model is trained on MSP-Podcasts dataset \cite{Lotfian2019}, a large affective speech corpus derived from YouTube with valence, arousal, and dominance scores. For this finetuning, the combined samples on the MSP-Podcasts dataset have a combined length of roughly 21 hours. The model extracted the speech embedding from the dataset (HUME-VB) with a size of 1024 dimensions for each utterance. The output layers of the model, i.e., the logits, are the scores of arousal, dominance, and valence, in ranges [0, 1]. We experimented with two variants of speech embedding with this model. The first embedding is the hidden states of the last layer before the output layer (1024-dim), and the second embedding is the concatenation of hidden states with the logits (1027-dim). We named the first embedding ``w2v2-R-er" (wav2vec 2.0 robust emotion recognition), and the second embedding``w2v2-R-vad" (wav2vec 2.0 robust emotion recognition with valence, arousal, and dominance). Notice the term of acoustic embedding (extracted from vocal bursts) here is used instead of (traditonal) acoustic features or speech embedding (extracted from speech).

\subsection{Classifier}
\begin{figure*}[htbp]
  \centering
  \includegraphics[width=0.85\textwidth]{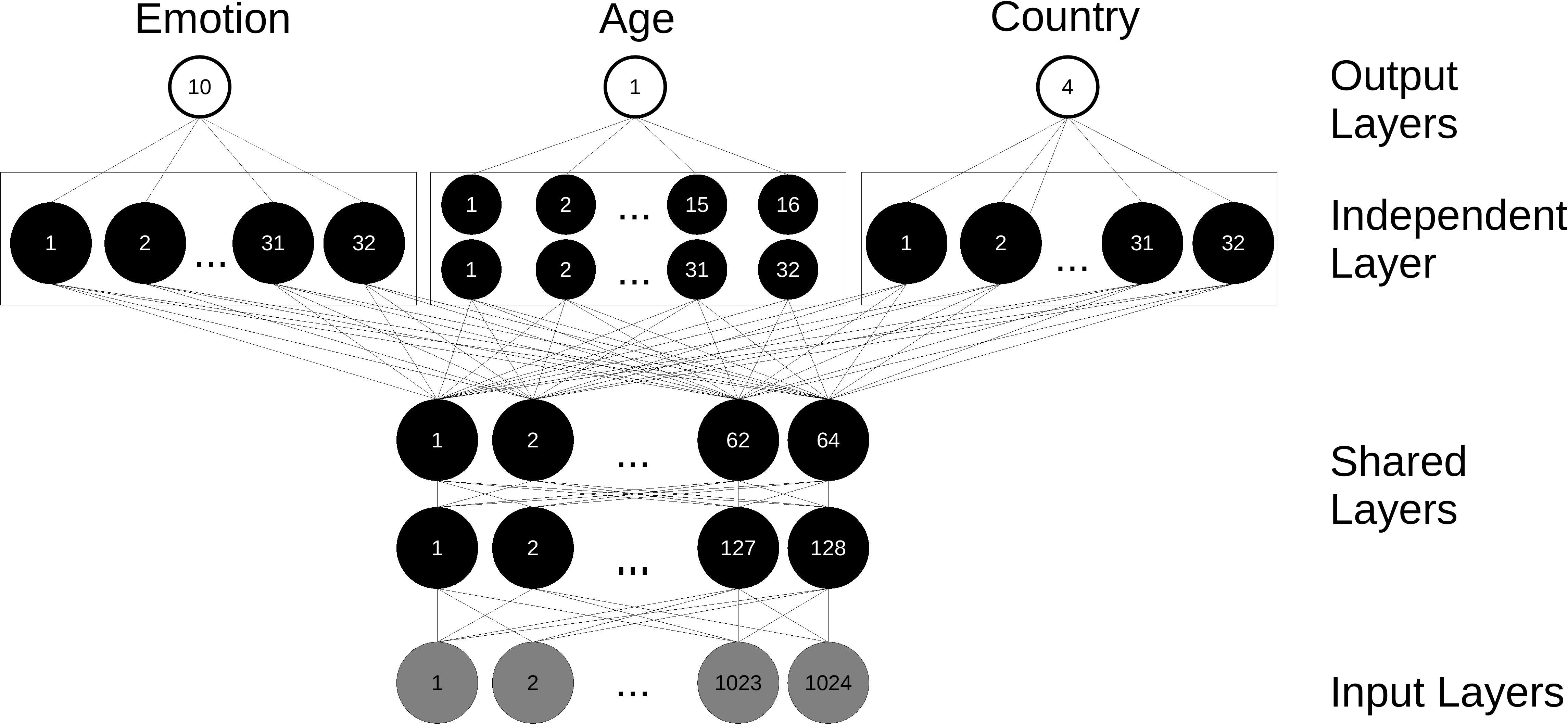}
  \caption{Architecture of multitask learning (MTL) for predicting emotion, age, and country. Emotion and age tasks are regressions; country prediction is a classification task; the number within the circle denotes units/nodes.}
  \label{fig:arch}
\end{figure*}

We employed a model of multitask learning adopted from \cite{Baird2022}. The architecture of the model is shown in Figure \ref{fig:arch}. The model accepts the input of acoustic features depending on the size of the features. For instance, the dimension for w2v2-R-er is 1024. Then, two linear layers are stacked as shared layers. The number of nodes for each layer is 128 and 64, respectively. We applied layer normalization \cite{Ba2015} for each linear layer, followed by LeakyReLU activation functions. A group of layers builds up the independent layers for each task. For emotion and country prediction, there is only a layer followed by output layers. For age prediction, we used two independent layers to bridge the gap in the big number of nodes from the shared layer (64) to the single-node output layer. The nodes for these independent layers are 32 and 16, respectively.

The loss function minimizes losses of three tasks: MSE for emotion recognition age prediction, cross-entropy (CE) for country prediction. The total loss function ($\mathcal{L}_T$) is weighting sum of three losses given by the following formula,

\begin{equation}
  \mathcal{L}_T = \sum_{i=1}^3 \left( \frac{\mathcal{L}_i}{2e^{\alpha}} + \frac{\alpha}{2}  \right).
  % \mathcal{L}=\left[\frac{1}{2e^{\alpha}} MSE_{emo}+\frac{\alpha}{2} \right]+\left[\frac{1}{2 e^{\beta}} \text { CE_{cou}}+\frac{\beta}{2} \right]+\left[\frac{1}{2 e^{\gamma}} MSE_{\text {age}}+\frac{\gamma}{2} \right]
\end{equation}
The coefficient of $\alpha$ is set to be 0.34, 0.33, and 0.33 for emotion, country, and age, respectively.

\subsection{Evaluation Metrics}
The evaluation of three paralinguistic tasks in this study used specific metric for each task. The emotion recognition is evaluated in concordance correlation coefficient (CCC), age recognition is evaluated in mean absolute error (MAE), and nationality prediction is evaluated in unweighted average recall (UAR). These metrics are described below,

\begin{equation}
  \mathcal{CCC}=\frac{2 \sigma_{x y}^{2}}{\sigma_{x}^{2}+\sigma_{y}^{2}+\left(\mu_{x}-\mu_{y}\right)^{2}}.
\end{equation}
The average CCC scores for ten emotion categories is the mean values,
\begin{equation}
  \hat{\mathcal{C}}=\sum_{i=1}^{10} \mathcal{CCC}_{i} / 10.
\end{equation}

CCC is in range [-1, 1] with -1 for perfect disconcordance, 0 for absence of concordance/disconcordance, and 1 for perfect concordance.

Next is metric to evaluate the performance of nationality/country prediction that is unweighted accuracy, also known as unweighted average recall (UAR) and balanced accuracy. Unweighted accuracy is formulated as,

\begin{equation}
  \hat{\mathcal{U}} = \frac{1}{4} \sum_{i=1}^{4}Recall_i
\end{equation}

\noindent
where $i$ is the corresponding country class, and $4$ is the number of countries (USA, China, South Africa, and Venezuela). UAR ranges in 0-100 in \% or 0-1 in normalized score.

The metric to evaluate the last task, age prediction, is mean absolute error (MAE). MAE is a common metric for evaluating regression, and it is scale-dependent. The lower the scores, the better the age prediction. The formulation of MAE is given by

\begin{equation}
  MAE=\frac{1}{n} \sum_{i=1}^{n}\left|x_{i}-y_{i}\right|,
\end{equation}
where $n$ is the number of samples in the evaluation or test sets (which one is used to calculate the score).

Since MAE is scale-dependent, we inverted the MAE scores for consistency with the previous two metrics,
\begin{equation}
  \hat{\mathcal{M}}=1 / MAE.
\end{equation}

Now, for all metrics (CCC, UAR, 1/MAE), the higher scores, the better predictions of emotion, country, and age. Finally, for calculating overall performance, we used the harmonic mean of three metrics above \cite{Baird2022},
\begin{equation}
  S_{MTL} = \dfrac{3}{(1 / \hat{\mathcal{C} } + 1 / \hat{\mathcal{M} } + 1/ \hat{\mathcal{U} })}.
\end{equation}

$S_{MTL}$ is our main metric to judge the performance of the evaluated methods (hyperparameters, features, normalization). Other previous metrics are used to determine the performance of the corresponding method for individual tasks in multitask learning.

\section{Results and Discussion}
We presented our results in different ablation studies: choosing the right seed, comparing different acoustic features, effects of normalization, and test results. For each study, we run the experiment five times for each setting. For instance, in choosing the right seed, we run the experiment five times on seed "101". The reported results are the average scores, except for comparing acoustic features. The reported scores for w2v2-R-er and w2v2-R-vad are chosen from the best from five different runs, similar to the baseline. For the baseline results, we quote the scores from the source \cite{Baird2022}.

\subsection{Choosing the right seed}
Seed initialization is an important step in deep learning methods \cite{Pepino2020,Macary2020}. Here, as a first step, we choose the seven different seed values to choose the best one. We evaluated seed values 42, 101, 102, 103, 104, 105, 106. The reason for adding seed 42 to the other six seed values given in the baseline is due to its common use in the deep learning community.

Table \ref{tab:seed} shows the results of using different seed values for multitask learning of emotion, age, and country on the validation set. The results are average scores of five trials with their standard deviation. Since the values of STD are similar in other ablation tests in this study, we only report these STD scores in Table \ref{tab:seed}. The following reported scores, except stated, used the seed values of "106" which obtained the highest $S_{MTL}$ score from seed values evaluation.

\begin{table*}[htbp]
  \caption{Evaluation of different seed values for initialization on validation set (mean values $\pm$ standard deviation); Feature: w2v2-R-er}
  \centering\begin{tabular}{c c c c c}
    \hline
    Seed value & Emo-CCC & Cou-UAR & Age-1/MAE & $S_{MTL}$ \\
    \hline
    42	& 0.534	$\pm$ 0.010	& 0.520	$\pm$ 0.006	& 0.245	$\pm$ 0.005	& 0.3806	$\pm$ 0.003 \\
    101	& 0.543	$\pm$ 0.006	& 0.520	$\pm$ 0.004	& 0.242	$\pm$ 0.003	& 0.3798	$\pm$ 0.004 \\
    102	& 0.529	$\pm$ 0.002	& 0.531	$\pm$ 0.003	& 0.239	$\pm$ 0.002	& 0.3773	$\pm$ 0.002 \\
    103	& 0.539	$\pm$ 0.004	& 0.512	$\pm$ 0.005	& 0.236	$\pm$ 0.003	& 0.3728	$\pm$ 0.003 \\
    104	& 0.537	$\pm$ 0.004	& 0.527	$\pm$ 0.006	& 0.247	$\pm$ 0.005	& 0.3838	$\pm$ 0.004 \\
    105	& 0.541	$\pm$ 0.005	& 0.518	$\pm$ 0.006	& 0.234	$\pm$ 0.008	& 0.3725	$\pm$ 0.008 \\
    106	& 0.537	$\pm$ 0.010	& 0.527	$\pm$ 0.006	& 0.247	$\pm$ 0.005	& \textbf{0.3844	$\pm$ 0.003} \\
    \hline
  \end{tabular}
  \label{tab:seed}
\end{table*}

\subsection{Comparison of different features}
The main focus of this study is to evaluate the acoustic embedding extracted using pre-trained models trained on the emotional speech dataset. Most speech processing tasks, especially speech emotion recognition, are trained using handcrafted acoustic feature extractors (e.g., MFCC, spectrogram, or mel-spectrogram) \cite{Atmaja2022}. Others are using pre-trained models but trained on neutral speech (e.g., wav2vec2 2.0, HuBERT, WavLM). In this study, we utilized a pre-trained model \cite{Wagner2022a,Wagner2022a} built using wav2vec 2.0 Robust on affective speech dataset.

Table \ref{tab:feat} proves our presumption that our model will surpass the baseline scores. Both w2v2-R-er and w2v2-R-vad obtain higher $S_{MTL}$ scores than the baseline scores on the same configuration (batch size, seed, and other hyperparameters). Specifically, these scores obtained by two acoustic embeddings show the most remarkable improvement in emotion recognition score, in which the model to extract the acoustic embedding is trained. Not only for the emotion recognition task but both scores for age and country predictions were also improved. The pre-trained model is shown to be helpful on other tasks probably due to the similarity of the task (paralinguistics and non-linguistic tasks), and the data trained to build the model contains the age and country information embedded on the extracted acoustic embedding. The data to train the model is MSP-Podcast in the English language. Although the dataset contains English only language, the pre-trained model may be able to discriminate between English with non-English language (related to country prediction) as in anomaly detection problems.

%  std is given on different seed evaluation since other runs are similar.
\begin{table}[htbp]
  \caption{Comparison of best scores from different acoustic features on validation set; scores from ComParE to DeepSpec are obtained from \cite{Baird2022}}
  \centering\begin{tabular}{l c c c c c}
  \hline
  Feature & Dims. &  CCC & UAR & 1/MAE & $S_{MTL}$ \\
  \hline
  ComParE	& 6373 &0.416	& 0.506	& 0.237	& 0.349 \\
  eGeMAPS	& 88	 &0.353	& 0.423	& 0.249	& 0.324 \\
  BoAW	& 125	&   0.335	& 0.417	& 0.234	& 0.311 \\
  	    & 250	&   0.354	& 0.423	& 0.238	& 0.319 \\
  	    & 500	&   0.374	& 0.432	& 0.218	& 0.314 \\
  	    & 1000	& 0.384	& 0.438	& 0.225	& 0.321 \\
  DeepSpec	  & 4096 &  0.369	  & 0.456 	& 0.227	& 0.322 \\
  w2v2-R-er	  & 1024 &  0.533	& 0.523 & 	0.252 &	0.386 \\
  w2v2-R-vad	& 1027 &  0.534	  & 0.525 & 	0.253 &	\textbf{0.388} \\
  \hline
  \end{tabular}
  \label{tab:feat}
\end{table}

\subsection{Evaluation of different batch sizes}
We evaluated five different batch sizes since there is evidence the influence of batch size, particularly for emotion recognition task \cite{Wu2018}. As found in the \cite{Wu2018}, we also found that the smallest evaluated batch size in this study resulted in the best performance. As the batch size increases, the performance decreases. For the best performance on the use of batch size = 2 (scores in Table \ref{tab:batch} are average of 5 runs), we obtained the following scores: 0.534, 0.533, 0.267, and 0.401 for Emo-UAR, Age-CCC, Cou-1/MAE, and $S_{MTL}$, respectively. This is the highest $S_{MTL}$ score obtained in this study.

\begin{table}[htbp]
  \caption{Average scores of different batch sizes from 5 runs on validation set; Feature: w2v2-R-vad}
  \centering\begin{tabular}{c c c c c}
    \hline
    Batch size & CCC & UAR & 1/MAE & $S_{MTL}$ \\
    \hline
    2	  & 0.532	& 0.508	& 0.264	& \textbf{0.393} \\
    4	  & 0.535	& 0.502	& 0.241	& 0.374 \\
    8	  & 0.534	& 0.525	& 0.252	& 0.387 \\
    16	& 0.535	& 0.511	& 0.236	& 0.372 \\
    32	& 0.516	& 0.505	& 0.228	& 0.361 \\
  \hline
  \end{tabular}
  \label{tab:batch}
\end{table}

\subsection{Effect of waveform normalization}
Normalization may affect the performance of deep learning, particularly in speech processing. This effect is due to the model usually only working on the standard input and being prone to high deviation. It has been proved that such normalizations are effective for deep learning, e.g., batch normalization \cite{Ioffe2015}, group normalization \cite{Wu2018}, and layer normalization \cite{Ba2015}. Aside from layer normalization, we also evaluated the model with waveform normalization. In this case, we utilize librosa toolkit \cite{McFee2020} to normalize the audio array (the amplitude of waveform), i.e., the output is in the range [-1, 1]. Then, two acoustic embeddings are created with these normalizations, namely w2v2-R-er-norm and w2v2-R-vad-norm. The results are shown in Table \ref{tab:norm}.

As shown in Table \ref{tab:norm}, there are no such improvements by normalizing the waveform of speech. This finding may be explained by the fact that emotion, age, and gender are related to the loudness of the waveform. Hence, such normalization of the waveform will remove important information that discriminates these paralinguistics labels. The results for two acoustic emebddings, w2v2-R-er and w2v2-R-vad, are consistent with and without normalization, highlighting the unnecessary processing of normalizing the speech waveform.

\begin{table}[htbp]
  \caption{Comparison of the acoustic embeddings with and without normalizations on validation set}
  \centering\begin{tabular}{l c c c c c}
    \hline
    Feature & Batch size & CCC & UAR & 1/MAE & $S_{MTL}$ \\
    \hline
    w2v2-R-er	      & 8	& 0.537	& 0.527	& 0.247	& 0.384 \\
    w2v2-R-er-norm	& 8	& 0.542	& 0.524	& 0.242	& 0.381 \\
    w2v2-R-vad	    & 2	& 0.532	& 0.508	& 0.264	& \textbf{0.393} \\
    w2v2-R-vad-norm	& 2	& 0.528	& 0.519	& 0.260	& 0.391 \\
    \hline
  \end{tabular}
  \label{tab:norm}
\end{table}

\subsection{Test results}
We presented the results of our predictions from the previous validation evaluation for the closed test set evaluation. Note that the best result from the validation set reported in this study was not submitted to obtain the test scores due to time limitations (for obtaining the test scores as part of The ICML 2022 Expressive Vocalizations Workshop and Competition). In the submitted results, we evaluated two predictions from both w2v2-R-er features set with different seed and batch sizes. The first prediction is with a batch size of 8 and a seed of 42. The second prediction is with a batch size of 4 and a seed of 106. Note that this last prediction is submitted with the original architecture \cite{Baird2022} with one hidden layer for all independent layers.

Table \ref{tab:test} shows the score of harmonic mean ($S_{MTL}$) from our predictions (last two rows) and the baseline \cite{Baird2022}. Although we did not use the best-reported validation scores in this study, our submitted results are still higher than the baseline results. These scores reveal the effectiveness of the evaluated acoustic embedding, w2v2-R-er, which were extracted using the robust version of wav2vec 2.0 \cite{Baevski2020a} on the MSP-Podcast emotional speech dataset \cite{Lotfian2019}. We believe that the test score will be even higher for w2v2-R-vad with batch size 2 since there are remarkable improvements on the validation set by using this configuration (the best validation $S_{MTL} = 0.401$).

\begin{table}[htbp]
\caption{Results of the submitted predictions on the test set; 'best' in seed column is obtained from values in range [101, 102, 103, 104, 105, 106].}
\centering\begin{tabular}{l c c c}
\hline
  Feature & Batch size & Seed & $S_{MTL}$ \\
\hline
ComParE	  & 8	& best	& 0.335 \\
eGeMAPS	  & 8	& best	& 0.214 \\
BoAW-125	& 8	& best	& 0.299 \\
BoAW-250	& 8	& best	& 0.305 \\
BoAW-500	& 8	& best	& 0.302 \\
BoAW-1000	& 8	& best	& 0.307 \\
BoAW-2000	& 8	& best	& 0.303 \\
DeepSpec	& 8	& best	& 0.305 \\
w2v2-R-er	& 8	& 42	  & 0.358 \\
w2v2-R-er	& 4	& 106	 & \textbf{0.378} \\
\hline
  \end{tabular}
  \label{tab:test}
\end{table}
\section{Conclusions}
In this paper, we reported evaluations of multitask learning to jointly predict emotion, age, and country by using acoustic emebdding extracted from a pre-trained model. The model is finetuned on an affective speech dataset. The extracted acoustic emebddings were fed to an architecture consisting of shared layers and independent layers for these three tasks. The results showed improvements over the baseline methods with the common speech representations (ComParE, eGeMAPS, BoAW, and DeepSpectrum). Two variants of acoustic embeddings are evaluated with original hidden states and concatenation of hidden states with logits. The latter performed best on the validation set. We conducted ablation studies on different seeds, batch sizes, and normalization. The study finds the optimum seed and batch size of the evaluated ranges and finds no improvement in performing waveform normalizations. While this study treated all emotions in the same weights, future studies may be directed to adjust these weights for optimum emotion recognition, as well as to improve overall harmonic mean evaluation for all tasks in Multitask learning.

\section*{Acknowledgment}
This paper is based on results obtained from a project, JPNP20006, commissioned by the New Energy and Industrial Technology Development Organization (NEDO), Japan.

\bibliography{exvo}

\begin{thebibliography}{10}
\providecommand{\url}[1]{#1}
\csname url@samestyle\endcsname
\providecommand{\newblock}{\relax}
\providecommand{\bibinfo}[2]{#2}
\providecommand{\BIBentrySTDinterwordspacing}{\spaceskip=0pt\relax}
\providecommand{\BIBentryALTinterwordstretchfactor}{4}
\providecommand{\BIBentryALTinterwordspacing}{\spaceskip=\fontdimen2\font plus
\BIBentryALTinterwordstretchfactor\fontdimen3\font minus
  \fontdimen4\font\relax}
\providecommand{\BIBforeignlanguage}[2]{{%
\expandafter\ifx\csname l@#1\endcsname\relax
\typeout{** WARNING: IEEEtran.bst: No hyphenation pattern has been}%
\typeout{** loaded for the language `#1'. Using the pattern for}%
\typeout{** the default language instead.}%
\else
\language=\csname l@#1\endcsname
\fi
#2}}
\providecommand{\BIBdecl}{\relax}
\BIBdecl

\bibitem{Batliner2020}
A.~Batliner, S.~Hantke, and B.~W. Schuller, ``{Ethics and Good Practice in
  Computational Paralinguistics},'' \emph{IEEE Trans. Affect. Comput.}, vol.
  3045, no.~c, pp. 1--1, 2020.

\bibitem{Baird2022}
A.~Baird, P.~Tzirakis, G.~Gidel, M.~Jiralerspong, E.~B. Muller, K.~Mathewson,
  B.~Schuller, E.~Cambria, D.~Keltner, and A.~Cowen, ``{The ICML 2022
  Expressive Vocalizations Workshop and Competition: Recognizing, Generating,
  and Personalizing Vocal Bursts},'' 2022.

\bibitem{Cowen2019}
A.~S. Cowen, H.~A. Elfenbein, P.~Laukka, and D.~Keltner, ``{Mapping 24 emotions
  conveyed by brief human vocalization.}'' \emph{Am. Psychol.}, vol.~74, no.~6,
  pp. 698--712, sep 2019.

\bibitem{Scherer1986}
K.~R. Scherer, ``\BIBforeignlanguage{eng}{{Vocal affect expression: a review
  and a model for future research.}}'' \emph{\BIBforeignlanguage{eng}{Psychol.
  Bull.}}, vol.~99, no.~2, pp. 143--165, mar 1986.

\bibitem{Schuller2022}
B.~W. Schuller, A.~Batliner, S.~Amiriparian, C.~Bergler, M.~Gerczuk, N.~Holz,
  P.~Larrouy-Maestri, S.~P. Bayerl, K.~Riedhammer, A.~Mallol-Ragolta,
  M.~Pateraki, H.~Coppock, I.~Kiskin, M.~Sinka, and S.~Roberts, ``{The ACM
  Multimedia 2022 Computational Paralinguistics Challenge: Vocalisations,
  Stuttering, Activity, {\&} Mosquitoes},'' 2022.

\bibitem{Kaya2017}
\BIBentryALTinterwordspacing
H.~Kaya, A.~A. Salah, A.~Karpov, O.~Frolova, A.~Grigorev, and E.~Lyakso,
  ``{Emotion, age, and gender classification in children's speech by humans and
  machines},'' \emph{Comput. Speech Lang.}, vol.~46, pp. 268--283, 2017.
  [Online]. Available: \url{http://dx.doi.org/10.1016/j.csl.2017.06.002}
\BIBentrySTDinterwordspacing

\bibitem{Zhao2018a}
\BIBentryALTinterwordspacing
H.~Zhao, N.~Ye, and R.~Wang, ``{Transferring Age and Gender Attributes for
  Dimensional Emotion Prediction from Big Speech Data Using Hierarchical Deep
  Learning},'' in \emph{2018 IEEE 4th Int. Conf. Big Data Secur. Cloud
  (BigDataSecurity), IEEE Int. Conf. High Perform. Smart Comput. IEEE Int.
  Conf. Intell. Data Secur.}\hskip 1em plus 0.5em minus 0.4em\relax IEEE, may
  2018, pp. 20--24. [Online]. Available:
  \url{https://ieeexplore.ieee.org/document/8552276/}
\BIBentrySTDinterwordspacing

\bibitem{parthasarathy2017jointly}
S.~Parthasarathy and C.~Busso, ``{Jointly Predicting Arousal, Valence and
  Dominance with Multi-Task Learning},'' in \emph{Interspeech 2017}.\hskip 1em
  plus 0.5em minus 0.4em\relax ISCA: ISCA, aug 2017, pp. 1103--1107.

\bibitem{Lee2019}
S.-w. Lee, ``{The Generalization Effect for Multilingual Speech Emotion
  Recognition across Heterogeneous Languages},'' in \emph{ICASSP 2019 - 2019
  IEEE Int. Conf. Acoust. Speech Signal Process.}\hskip 1em plus 0.5em minus
  0.4em\relax IEEE, may 2019, pp. 5881--5885.

\bibitem{Kim2017}
\BIBentryALTinterwordspacing
J.~Kim, G.~Englebienne, K.~P. Truong, and V.~Evers, ``{Towards Speech Emotion
  Recognition “in the Wild” Using Aggregated Corpora and Deep Multi-Task
  Learning},'' in \emph{Interspeech 2017}.\hskip 1em plus 0.5em minus
  0.4em\relax ISCA, aug 2017, pp. 1113--1117. [Online]. Available:
  \url{http://www.isca-speech.org/archive/Interspeech{\_}2017/abstracts/0736.html}
\BIBentrySTDinterwordspacing

\bibitem{Atmaja2020b}
\BIBentryALTinterwordspacing
B.~T. Atmaja and M.~Akagi, ``{Evaluation of Error and Correlation-Based Loss
  Functions For Multitask Learning Dimensional Speech Emotion Recognition},''
  \emph{J. Phys. Conf. Ser.}, vol. 1896, no.~1, p. 012004, mar 2020. [Online].
  Available: \url{http://arxiv.org/abs/2003.10724}
\BIBentrySTDinterwordspacing

\bibitem{Li2020}
K.~Li, M.~Akagi, Y.~Wu, and J.~Dang, ``{Segment-level Effects of Gender ,
  Nationality and Emotion Information on Text-independent Speaker
  Verification},'' in \emph{Interspeech}, 2020.

\bibitem{Cai2021}
X.~Cai, J.~Yuan, R.~Zheng, L.~Huang, and K.~Church, ``{Speech Emotion
  Recognition with Multi-Task Learning},'' in \emph{Interspeech 2021}.\hskip
  1em plus 0.5em minus 0.4em\relax ISCA: ISCA, aug 2021, pp. 4508--4512.

\bibitem{Atmaja2022a}
B.~T. Atmaja, A.~Sasou, and M.~Akagi, ``{Speech Emotion and Naturalness
  Recognitions with Multitask and Single-task Learnings},'' \emph{IEEE Access},
  pp. 1--1, 2022.

\bibitem{Wagner2022}
\BIBentryALTinterwordspacing
B.~W. {Wagner, Johannes, Triantafyllopoulos, Andreas, Wierstorf, Hagen,
  Schmitt, Maximilian, Eyben, Florian, Schuller}, ``{Model for Dimensional
  Speech Emotion Recognition based on Wav2vec 2.0 (1.1.0)}.'' [Online].
  Available: \url{https://doi.org/10.5281/zenodo.6221127}
\BIBentrySTDinterwordspacing

\bibitem{Wagner2022a}
\BIBentryALTinterwordspacing
J.~Wagner, A.~Triantafyllopoulos, H.~Wierstorf, M.~Schmitt, F.~Burkhardt,
  F.~Eyben, and B.~W. Schuller, ``{Dawn of the transformer era in speech
  emotion recognition: closing the valence gap},'' 2022. [Online]. Available:
  \url{http://arxiv.org/abs/2203.07378}
\BIBentrySTDinterwordspacing

\bibitem{Hsu2021a}
W.~N. Hsu, A.~Sriram, A.~Baevski, T.~Likhomanenko, Q.~Xu, V.~Pratap, J.~Kahn,
  A.~Lee, R.~Collobert, G.~Synnaeve, and M.~Auli, ``{Robust wav2vec 2.0:
  Analyzing domain shift in self-supervised pre-training},'' in \emph{Proc.
  Annu. Conf. Int. Speech Commun. Assoc. INTERSPEECH}, vol.~3, 2021, pp.
  2123--2127.

\bibitem{Lotfian2019}
R.~Lotfian and C.~Busso, ``{Building Naturalistic Emotionally Balanced Speech
  Corpus by Retrieving Emotional Speech from Existing Podcast Recordings},''
  \emph{IEEE Trans. Affect. Comput.}, vol.~10, no.~4, pp. 471--483, 2019.

\bibitem{Ba2015}
J.~L. Ba, J.~R. Kiros, and G.~E. Hinton, ``{Layer Normalization},''
  \emph{arXiv:1607.06450v1}, 2015.

\bibitem{Pepino2020}
\BIBentryALTinterwordspacing
L.~Pepino, P.~Riera, L.~Ferrer, and A.~Gravano, ``{Fusion Approaches for
  Emotion Recognition from Speech Using Acoustic and Text-Based Features},'' in
  \emph{ICASSP 2020 - 2020 IEEE Int. Conf. Acoust. Speech Signal
  Process.}\hskip 1em plus 0.5em minus 0.4em\relax IEEE, may 2020, pp.
  6484--6488. [Online]. Available:
  \url{https://ieeexplore.ieee.org/document/9054709/}
\BIBentrySTDinterwordspacing

\bibitem{Macary2020}
M.~Macary, M.~Lebourdais, M.~Tahon, Y.~Est{\`{e}}ve, and A.~Rousseau,
  ``{Multi-corpus Experiment on Continuous Speech Emotion Recognition:
  Convolution or Recurrence?}'' in \emph{Int. Conf. Speech Comput.}, 2020, pp.
  304--314.

\bibitem{Atmaja2022}
B.~T. Atmaja, A.~Sasou, and M.~Akagi, ``{Survey on bimodal speech emotion
  recognition from acoustic and linguistic information fusion},'' \emph{Speech
  Commun.}, vol. 140, pp. 11--28, may 2022.

\bibitem{Wu2018}
\BIBentryALTinterwordspacing
Y.~Wu and K.~He, ``{Group Normalization},'' 2018. [Online]. Available:
  \url{https://github.com/facebookresearch/Detectron/}
\BIBentrySTDinterwordspacing

\bibitem{Ioffe2015}
S.~Ioffe and C.~Szegedy, ``{Batch normalization: Accelerating deep network
  training by reducing internal covariate shift},'' in \emph{32nd Int. Conf.
  Mach. Learn. ICML 2015}, vol.~1, 2015, pp. 448--456.

\bibitem{McFee2020}
\BIBentryALTinterwordspacing
B.~McFee, V.~Lostanlen, M.~McVicar, A.~Metsai, S.~Balke, C.~Thom{\'{e}},
  C.~Raffel, A.~Malek, D.~Lee, F.~Zalkow, K.~Lee, O.~Nieto, J.~Mason, D.~Ellis,
  R.~Yamamoto, S.~Seyfarth, E.~Battenberg, V.~Morozov, R.~Bittner, K.~Choi,
  J.~Moore, Z.~Wei, S.~Hidaka, Nullmightybofo, P.~Friesch, F.-R. St{\"{o}}ter,
  D.~Here{\~{n}}{\'{u}}, T.~Kim, M.~Vollrath, and A.~Weiss, ``librosa/librosa:
  0.7.2,'' jan 2020. [Online]. Available:
  \url{https://zenodo.org/record/3606573}
\BIBentrySTDinterwordspacing

\bibitem{Baevski2020a}
A.~Baevski, H.~Zhou, A.~Mohamed, and M.~Auli, ``{wav2vec 2.0: A framework for
  self-supervised learning of speech representations},'' \emph{Adv. Neural Inf.
  Process. Syst.}, vol. 2020-Decem, no. Figure 1, pp. 1--12, 2020.

\end{thebibliography}
\bibliographystyle{IEEEtran}

\end{document}